\documentclass[sigplan, nonacm]{acmart}
\newcommand{\ncite}[1]{}

\newcommand{\done}[1]{}

\usepackage{booktabs}

\usepackage{xspace}
\usepackage{graphicx}
\usepackage{caption}
\usepackage{subcaption}
\usepackage{multicol}

\begin{document}

\title{Confidential Deep Learning: Executing Proprietary Models on Untrusted  Devices} 

\author{Peter M. VanNostrand$^\ast$ \quad Ioannis Kyriazis$^\dagger$ \quad Michelle Cheng$^+$ \quad Tian Guo$^\dagger$ \quad Robert J. Walls$^\dagger$}
\affiliation{
    \vspace{-2mm}
    \begin{center}
    \begin{multicols}{3}
        University at Buffalo$^\ast$ \\ pmvannos@buffalo.edu$^\ast$ \vfill\null\columnbreak
        Colby College$^+$ \\ michelle.cheng@colby.edu$^+$ \vfill\null\columnbreak
        Worcester Polytechnic Institute$^\dagger$ \\ {ikyriazis,tian,rjwalls}@wpi.edu$^\dagger$ \vfill\null\columnbreak
    \end{multicols}
    \end{center}
    \vspace{-6mm}
}

\renewcommand{\shortauthors}{VanNostrand et al.}

\begin{abstract}
	Performing deep learning on end-user devices provides fast
	offline inference results and can help protect the user's privacy.
	However, running models on untrusted client devices reveals model
	information which may be proprietary, i.e., the operating system or
	other applications on end-user devices may be manipulated to copy and
	redistribute this information, infringing on the model provider's
	intellectual property. We propose the use of ARM TrustZone, a
	hardware-based security feature present in most phones, to
	confidentially run a proprietary model on an untrusted end-user device.
	We explore the limitations and design challenges of using TrustZone and
	examine potential approaches for confidential deep learning within this
	environment. Of particular interest is providing robust protection of
proprietary model information while minimizing total performance overhead.
\end{abstract}

\ccsdesc[500]{Security and privacy~Hardware-based security protocols}
\ccsdesc[300]{Computing methodologies~Neural networks}

\keywords{ARM TrustZone, Hardware Security, Convolutional Neural Networks}

\maketitle

\section{Introduction}
\label{sec:intro}

Mobile devices have traditionally adopted a client-server paradigm for
deep-learning-based inference, wherein the performance of the system is heavily
contingent on the network conditions~\done{GrandSLA, GuoIC2E, MoDI}. For
example, inference requests might require the frequent transfer of relatively
large quantities of data from the mobile device to the cloud server hosting the
Deep Neural Network (DNN) model---such is the case for speech or image
recognition. In the worst case, when the device lacks an internet connection
entirely, the user cannot benefit from these inference services at all.
One promising solution to these network problems is to execute the DNN models
directly on the user's device.\done{see if we can find a peer-reviewed paper.}

However, model execution on end-user devices presents its own novel challenges
for the model providers. In this work, we focus on one such challenge, namely,
\emph{maintaining the confidentiality of a proprietary model running on an
untrusted end-user device}. In general, a model may be considered intellectual
property when it adopts a custom architecture or when it was trained using
proprietary data. As users may have full access to the hardware and software of
their devices, the client operating system or other applications may be
manipulated to copy and redistribute proprietary models, infringing on the
model provider's intellectual property. We refer to this as the
\emph{confidential deep learning} problem.

In this paper, we demonstrate how a \emph{Trusted Execution Environment (TEE)}
can serve as the foundation for confidential deep learning on untrusted
devices. Both x86 and ARM architectures already provide hardware mechanisms
that support trusted execution environments. These mechanisms---SGX on
x86~\cite{mckeen2013innovative} and TrustZone on ARM~\cite{TrustZoneDocs}---are
widely deployed but rarely used by applications. While the technical details
vary between architecture and TEE implementation, TEEs typically provide three
fundamental capabilities that providers can leverage to keep models 
confidential. First, TEEs provide an isolated environment for code
execution that cannot be manipulated by other software running on the
device, including the untrusted operating system. Second, TEEs allow for a
model provider to remotely guarantee the integrity of the code loaded into the
TEE via remote attestation. Third, TEEs provide mechanisms to ensure that data
used inside of the environment cannot be read or manipulated externally and
that the data never leaves the TEE without being encrypted.  

However, a number of significant research challenges prevent model providers
from simply porting existing DNN inference code. First, operations performed in
a trusted execution environment typically run orders of magnitude slower than if
the same operations were performed in the untrusted
environment~\cite{tramer2018slalom}. Similarly, the TEE does not provide code
with the same abstractions (e.g., system calls, dynamic threading) or
capabilities (e.g., memory is limited and page swapping is expensive) as normal
code. Consequently, the inference code, and perhaps the models as well, must
be heavily modified to run within the TEE. In short, both the performance and
engineering costs are high.

We explore the performance tradeoffs and design challenges of
confidential deep learning in trusted execution environments. We focus
exclusively on ARM devices, as ARM is the dominant architecture for
end-user devices such as mobile phones. To provide a realistic context, we
consider existing hardware, TrustZone~\cite{TrustZoneDocs}; an existing TEE
implementation, OP-TEE~\cite{OP-TEE}; and an existing inference framework, 
DarkNet~\cite{darknet13}.

 \section{Background}
\label{sec:background}

\subsection{ARM TrustZone}
\label{sec:arm_trustzone}

\emph{ARM TrustZone} is a hardware security module present in most mobile
phones~\cite{TrustZoneDocs}. The core principle of TrustZone is the isolation
of secure and non-secure operations. To achieve this, TrustZone implements two
independent execution environments, called the \emph{Secure world} and
\emph{Normal world}, that run simultaneously on the same processor core.
TrustZone divides physical memory and peripherals between these two worlds such
that processes executing in the Normal world are only able to access a subset
of memory and peripherals. It also provides mechanisms to securely context
switch between the worlds. As the phone's operating system is placed in the
Normal world, TrustZone can be used as the foundation for a trusted execution
environment.

\subsection{Trusted Execution Environments} 

A \textit{trusted execution environment} (TEE) is an isolated environment that
runs in parallel with the Normal world OS. TEE implementations range from
simple libraries to full OS kernels and while the exact capabilities vary, a
TEE implementation must guarantee the confidentiality and integrity of the code
and data located inside the TEE. In this paper, we focus on OP-TEE, a trusted
execution environment built on TrustZone that follows the GlobalPlatform
standards~\cite{OP-TEE}\ncite{2a}.

Under OP-TEE, \textit{trusted applications} run in the Secure world (i.e., the
trusted execution environment), protected by use of TrustZone-provided
isolation and managed by an OP-TEE kernel. This kernel, which also runs in the
Secure world, provides useful APIs to trusted applications and performs all
interactions with the Normal world. As the kernel is designed to be lightweight,
it relies on the Normal world OS for various operations, including most system
calls, process scheduling, and reading files from storage. Each trusted
application is signed with a private pre-generated 2048-bit RSA key which can
be placed in a separate hardware security module to provide code signing.
OP-TEE will not execute trusted applications that fail the integrity
check\ncite{3a}.

Trusted applications, in essence, act as services that are called by \textit{client applications} running in the Normal world under the normal operating system. A \textit{context} connects the client application to the TEE, then a \textit{session} connects the client application to a specific trusted application inside the TEE. Once a session is established, the client application opens a communication channel to the trusted application. Data is passed between the client and trusted applications through the use of \textit{shared memory buffers}, these are regions of memory allocated in the Normal world by the untrusted OS which are then registered with OP-TEE and mapped into the memory space of the trusted application. This allows large amounts of data to be passed into and out of the secure world, but this memory is not secure and therefore cannot be trusted to hold unencrypted confidential information.

Lastly, the specific security guarantees of OP-TEE depend on the exact hardware
of the device. For example, though officially supported by OP-TEE, the
Raspberry Pi 3 Model B supports some TrustZone operations but lacks the
TrustZone Address Space Controller, which is the peripheral responsible for
partitioning the Normal and Secure world memory. Further, the Raspberry Pi 3
also lacks the hardware modules needed to securely store keys for code signing.

\subsection{Convolutional Neural Networks}
\label{sec:cnns}

We focus our initial efforts on a popular type of DNN called
\emph{convolutional neural networks (CNNs)}. CNNs are feedforward neural
networks that contain \emph{neurons} that are organized in layers. CNNs only
require each layer to be processed during inference, unlike recurrent neural networks
that might require accessing the same layer multiple times, making CNNs good
candidates for understanding the fundamental challenges of deep learning in
trusted execution environments.

The process of using a pre-trained CNN model to perform image classification is
referred to as \emph{inference} or the \emph{forward pass}. A pre-trained CNN
consists of at least one input and one output layer, as well as a number of
hidden layers. The total number of layers in a CNN is often referred to as the
depth of the network---hence deep neural networks. In the case of image
classification, the input layer is initialized as a single vector using raw
image pixel values and the output layer is a vector corresponding to the
likelihood of each potential classification label. Each hidden layer in-between
consists of neurons that typically consist of a weight vector and a bias, both
learned during training. For each neuron, we calculate an output value using
the outputs of the previous layer; specifically, we compute the dot product
between the neuron's input vector and its weight vector and then offset by a
bias. The input vector is a subset of the neuron outputs from the previous
layer as defined by the connections between the layers. In
essence, one can think of the forward pass of each layer in terms of matrix
multiplication.

 \section{Problem Definition and Threat Model}
\label{sec:threat}

A pre-trained CNN model may be considered intellectual property, or
\emph{proprietary}, when it adopts a custom architecture---e.g., the
number/type of layers and connections between them---or when the model was
trained using proprietary data. Our objective is a TEE architecture that
guarantees the confidentiality of a proprietary deep learning model---e.g.,
weights, structure---even when that model is running on a device controlled by
an untrusted user. In our threat model, we assume the untrusted user has
control over the device's operating system as well as total access to the
device's memory and storage. However, we also assume the presence of a trusted
execution environment loaded into the Secure World, e.g., OP-TEE. We discuss
the challenges of using such an environment for deep learning in
Section~\ref{sec:partitioning}.  The untrusted client may attempt to leak
proprietary model information, and we assume that they will succeed if that
data is ever stored unencrypted in memory or storage accessible by the untrusted
Normal world operating system.  

We consider that the model originates from a software developer as part of an
application installed into local storage accessible to the untrusted OS and
that the developer intends to keep this model confidential from the untrusted
user. We assume existing mechanisms in the TEE will prevent the loading of the
developer's application if it has been modified---indeed, such functionality is
already implemented in OP-TEE and was discussed previously. However, the
untrusted user may attempt to manipulate software running in the TEE by
manipulating the execution or responses of normal world system calls made by
the TEE, i.e., Iago attacks~\cite{Checkoway:2013:IAW:2499368.2451145}. Further,
the user may try to load malicious applications into the TEE. The inclusion of
malicious trusted applications is a significant threat as users will want to
download and install third party applications, which may come from compromised
or malicious vendors.

In our analysis we do not consider hardware attacks, side-channel attacks such as execution timing, or direct modification of the trusted execution environment before boot. We also exclude application signing and distribution. No defense is provided against Denial of Service (DoS) attacks on trusted applications. Similarly, no protection is provided against trusted applications revealing their own data intentionally or through leaks in their public APIs.

 \section{Model Partitioning to Support Confidential Deep Learning}
\label{sec:partitioning}

Below we walk through the design of a hypothetical framework for confidential
deep learning on untrusted devices. For this exercise, we assume a device with
TrustZone support using OP-TEE to provide the trusted execution environment.

\subsection{Motivation} \label{sec:motivation}

Imagine a theoretical machine learning framework tasked with performing
inference given an input (e.g., an image) and a pre-trained convolutional
neural network (CNN). Let us assume that the model is already present on the
device---saved to a file (or set of files) on disk---and will be loaded by the
framework as part of the inference process. Further, let's assume that all
inference calculations are done directly on the CPU, i.e., the system does not
have a GPU or other hardware accelerator that can be used for inference tasks. 

To perform inference using CNNs, our framework needs to calculate the
activation of many individual artificial neurons. To determine any given
neuron's activation, we need three things: the activation ($a$) of all the nodes
connected to it, a weight ($w$) for each connected node, and a bias ($b$) for
the given neuron. The neuron's output activation is a weighted sum of the input
neuron activations and the bias value $w_1a_1 + w_2a_2 + ... + w_na_n + b$. In
a fully connected layer, every neuron will depend on the same set of previous
activations, but will have a unique sets of weights and a unique bias. 

In conventional CNN execution, i.e., when memory is abundant and
confidentiality is not a concern, the weights and bias for every neuron in a
network are all pre-loaded into memory prior to neuron activation calculations.
This pre-loading enables neuron calculations to be performed efficiently using
matrix multiplication, which is highly optimized in modern CPUs. Additionally,
hardware accelerators such as GPUs allow many neurons in a given layer to be
computed simultaneously.

However, memory is not abundant inside of the trusted execution environment and
  pre-loading all of these values requires a substantial amount of memory. In
  the current version of OP-TEE, the physical memory dedicated to the secure
  world is statically configured at build time and trusted applications are
  limited to roughly 7MB of secure world memory. Compare this limit to the hundreds of MBs required
  to execute common CNN models---even MobileNet-v2
  which targets mobile device requires 630 MB of
  memory~\cite{bianco2018dnnsbench}.
  While redesigning
  OP-TEE to provide additional secure world memory is likely possible----for
  example, by modifying both operating systems to dynamically reallocate memory
  between worlds---the engineering challenges of implementing such an approach
  lead us to consider alternative solutions.

One way to counteract this memory limitation is to divide the neural network
  into partitions. Each partition would have its own weights file containing
  all the weights and biases for the neurons associated with that partition. By
  executing each partition separately, the entire model could be computed in
  pieces. If the partitions are selected to be sufficiently small, then the
  information necessary at runtime should be small enough to fit in the limited
  secure world memory. In the following sections we detail three potential
  methods to perform such partitioning along with the design constraints that
  motivate them.

\subsection{Layer-Based Partitioning} 

\begin{figure*} \begin{subfigure}{0.23\textwidth} \centering
\includegraphics[width=0.9\linewidth]{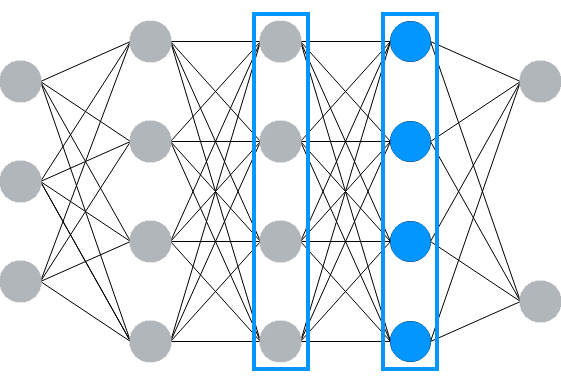}
\caption{Layer-Based Partitioning} \label{fig:partition_layered}
\end{subfigure} \hspace*{\fill} \begin{subfigure}{0.23\textwidth} \centering
\includegraphics[width=0.9\linewidth]{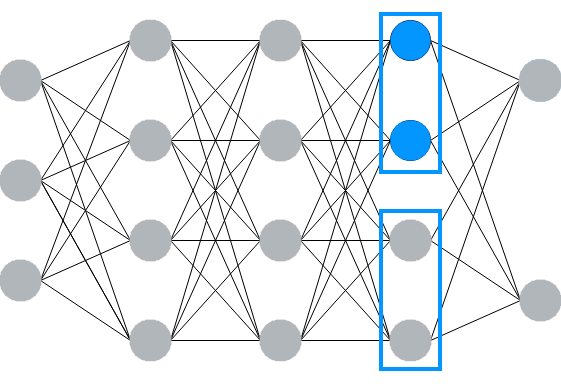}
\caption{Sub-Layer Partitioning} \label{fig:partition_sublayered}
\end{subfigure} \hspace*{\fill}
    \begin{subfigure}{0.22\textwidth}
    \includegraphics[width=0.9\linewidth]{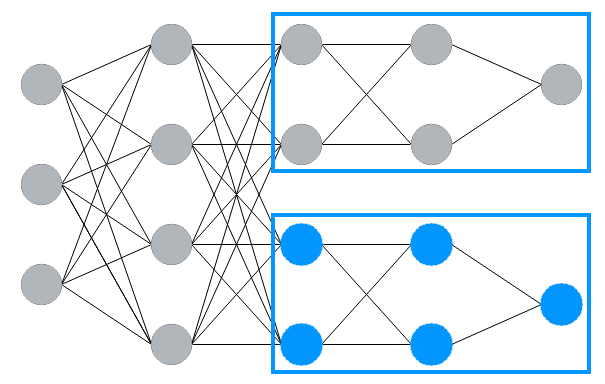}
    \caption{Branched Partitioning} \label{fig:partition_branched}
    \end{subfigure} \caption{Proposed Partition Schema for CNNs to Support Confidential Deep Learning} \label{fig:1}
    \end{figure*}

The simplest and most straightforward way to partition a CNN is to break it up
  by layer. As each layer depends only on the output of the previous layer,
  these divides act as natural partitions within existing network
  architectures. In this approach, each layer would be an independent partition
  as shown in Figure~\ref{fig:partition_layered} with its own set of weights
  and biases stored in a separate encrypted file. The encrypted file for one
  partition would be loaded into shared memory and then decrypted by the
  trusted application into secure memory. For the first partition, input data
  would be passed in from shared memory and the layer would be executed to
  compute the activation of all the neurons it contains. Once complete, the
  weights and biases of the current layer could be discarded and replaced with
  the appropriate values for the next layer. The current neuron activations
  would be stored in Secure world memory for use as inputs to the next layer.

This partitioning method is beneficial as the only values needed at a given
  time are the activations of the previous layer and the weights and biases for
  the current layer. This effectively limits the instantaneous memory footprint
  to that of a single layer, with the largest layer in the model determining
  the minimum amount of Secure world memory needed. For two fully connected
  layers each of $n$ neurons, this would be $n$ activation values, $n$ biases,
  and $n^2$ weights. However, if a single layer within a model has many
  neurons, it may still require a substantial amount of memory. While this may
  not occur for simple models, such as the MNIST LeNet classifier, we believe
  that many commonly used models such as MobileNet will encounter this issue
  due to their larger size. Overcoming this challenge requires further
  partitioning the model into smaller subsets.

\subsection{Sub-Layer Partitioning}

In layer-based partitioning, if any individual layer proves too large to run
entirely within the Secure world, then the model will fail to execute. This
limits the size of each layer to the size of the available memory and would
require model redesign to reduce the number of neurons in each layer,
potentially impacting model accuracy. 

To avoid redesigning the model, we propose sub-layer partitioning. In this
method the model is first partitioned by layer, and then further divided into
$p$ subsets of $s$ neurons where $1 < s \leq p$, this is shown in
Figure~\ref{fig:partition_sublayered} with $p = s = 2$. For two fully connected
layers each of $n$ neurons; $n$ activations, $sxn$ weights, and $s$ biases
would be needed at any given time. This method allows for a variable memory
footprint simply by adjusting the size of the subset given by $s$. At minimum
with $s = 1$, the model would be executed one neuron at a time with only the
previous $n$ activations, one set of $n$ weights, and one bias value being
required at any given time. This would likely reduce the needed memory by an
order of magnitude from simple layer based partitioning as $s$ could be
selected to be much smaller than $n$, with memory complexity growing as factor
of $sxn$ rather than $n^2$.  Note this also means that we cannot compute the
values of all neurons belonging to a layer at once using matrix multiplication,
leading to slower execution.  

For sub-layer partitioning, the defining characteristic of memory size is the
number of activations and weights that need to be stored from the previous
layer. While the current layer can easily be divided into small subsets, each
of those subsets will still require the full $n$ activations of the previous
layer. Should this require more memory than is available in the Secure world,
some of these activation values may need to be swapped into shared memory as
they are computed. These intermediate activation values represent proprietary
model information and would need to be encrypted before being moved into shared
memory. Unfortunately, this means that during the execution of one subset of
neurons, activation values would need to be loaded from shared memory and
decrypted into secure memory. This process would have to be repeated until all
activation values were accessed, meaning that the execution of each subset
would require several costly decryption calls. Worse yet, these values would
need to be cleared from memory to make room for the remaining activation
values,so every subset would have to re-decrypt the same information.

\subsection{Branched-Execution Partitioning}

To reduce the number of weights and activations required, and therefore reduce
the instantaneous memory footprint, we propose the use of branched execution
partitioning~\cite{5178994}. This method essentially splits the later layers of
CNNs into multiple mutually independent branches, as shown in
Figure~\ref{fig:partition_branched}. This reduces the number of intermediate
activation values and weights that would need to be stored in Secure memory as
the neurons of each branch depend only upon a portion of the neurons in the
previous layer. 

In this topology each branch is able to learn different features, and previous
work suggests the higher-level features extracted in the later layers of a
CNN tend to contain more significant information about the underlying training
data than the earlier layers~\cite{DBLP:journals/corr/abs-1907-06034}. As such,
one promising solution is to use a partially-branched model with the early,
unbranched, layers running in the Normal world and the later, branched, layers
running in the Secure world.

Partial-branching is especially beneficial from the perspective of memory usage
as it is common in CNNs for the early layers to contain the largest, highest
dimensional data. Executing the largest layers in the Normal world will allow
us to make use of abundant memory. The extracted features, which typically have
much smaller dimensionality than the input data, can be passed into the Secure
world to continue execution. In the Secure world, the branched nature of the
model will allow for sub-layer partitioning to be applied with a minimal number
of activations and weights needed at any given time. This should allow
activation values to be stored in memory and prevent the large encryption and
decryption overhead otherwise required.

\subsection{Summary} 

The limited available memory of trusted execution environments heavily
constrains the design of confidential deep learning frameworks. While model
partitioning is a promising mechanism to address these memory challenges,
designing performant partitioning schema is non-trivial and heavily influenced
by the structure of the underlying model. 

We believe that the three partitioning methods we propose are applicable under
different use cases. For small models, layer based partitioning seems promising
due to its simplicity and ability to run models without any modification to
their structures. For moderate-sized models, or models on platforms where
memory limitations are critical, sub-layer partitioning extends layer based
partitioning to further reduce memory usage.  However, as the size of layer
outputs approaches the available memory size, we risk significant performance
overhead. If model performance is paramount, branched-execution partitioning
has the potential to greatly decrease runtime, particularly for large models,
while still providing confidentiality given a tradeoff of additional
development time. 

 \section{Overhead of Layer-Based Partitioning}
\label{sec:results}

In this section, we briefly investigate the performance overhead introduced by
layer-based partitioning---the simplest method we propose. We identify two
factors that are likely to have the largest impact on performance: model
decryption and context switching.

We selected the Raspberry Pi 3 Model B, a hardware
platform that supports ARM TrustZone and implemented the hardware and software
components relevant to context switching and encryption on top of OP-TEE 3.5.0.
We used DarkNet~\cite{darknet13}, an open source neural network library, to
collect the baseline for CNN inference in the Normal world.  

When execution moves from the Normal
world to the Secure world (or vise versa), the hardware performs a context
switch, clearing the current program state and resuming execution in the other
world. This process requires resetting registers, switching
cache content, and clearing the execution pipeline~\cite{TrustZoneDocs}.
However, this overhead is only part of the total cost, as each transition
between a client application and a trusted application adds additional overhead
in the form of OP-TEE library code and OP-TEE secure OS operations. 
Using an instrumented client application, we measured an average context
switch time of \textbf{75.1 microseconds}.

As files are distributed in an encrypted form they
are protected from disclosure, but must be decrypted in the Secure world before
use. To estimate decryption overhead, we first measured the time to decrypt
chunks of memory of various sizes in a trusted application, finding an average 
cost of \textbf{163.7 nanoseconds} per block. 

Using DarkNet in the Normal world  and the MNIST LeNet model, we measured an
average inference time of \textbf{8.23 milliseconds}. With this baseline and
the costs of a context switch $t_{cs}$ and of decrypting a block $t_d$, we
estimate the overhead of layer-based partitioning as follows. The number of
context switches depends on the number of layers in the model. In this case,
LetNet has a total of 11 layers. The amount of information to be decrypted
depends on the model storage size. For the LeNet model, the weights file
contains 191,124 bytes and an additional 666 is needed for the model
configuration file, yielding a total of 191,790 bytes that need to be
decrypted. Hence, the additional execution time added by layer-based
partitioning is $t_{cs} * 11 * 2 + t_d * 191,790$, which comes out to be
33.05ms. Thus, even for simple models such as LeNet, we are looking at an
overhead of 4X. This suggests the importance of considering context switching
and decryption during both framework and model design.

 \section{Related Work}
\label{sec:related}

For context on trusted execution, we can look to
nascent research that focuses on using trusted execution environments, but in a
cloud server setting. Tramer and Boneh found a provider that can achieve lower
overhead when partitioning code if the primary goal is to ensure code integrity
rather than preserve the privacy of user information~\cite{tramer2018slalom}. This
observation naturally applies to the end-device scenario we are investigating, as user data remains private (without TEE protections) if it never leaves
the user's device. Yet, the work of Tramer and Boneh also leaves many open
questions. For instance, their work does not consider how to train the models
in the TEE, instead focusing solely on inferences. Further, their work
does not hide the model parameters from the untrusted OS and thus
proprietary model data could be leaked. Lastly, their framework was designed to 
work with Intel SGX rather than ARM TrustZone and thus needs to
be adapted for end-user devices.

Fan and Haddadi extended the use of trusted execution environments to develop
DarkNetP, an application for running DarkNet networks in ARM
TrustZone~\cite{DBLP:journals/corr/abs-1907-06034}. The focus of this work is on protecting the
privacy of data used to train neural networks by containing intermediate values
of the model within a TEE. As such, the authors provide an interesting
exploration of differential privacy techniques and explore how CNNs store
information about training data~\cite{DBLP:journals/corr/abs-1907-06034}. This
differs from our goal of ensuring the confidentiality of DNN models and
therefore Fan and Haddadi do not address the challenges of proprietary
information contained within model files nor the memory limitations of TEEs.

Given increasingly large DNN models and the need to
execute inference on resource constrained devices, researchers have proposed
model parallelism~\cite{model_parall} to speed up training and layer-based
partitioning~\cite{DBLP:journals/corr/abs-1907-06034,neurosurgeon,branchnets,5178994}
to speed up inference. Similarly, our formulation of layer-based partitioning
is a more generalized form of the DarkNetP approach~\cite{DBLP:journals/corr/abs-1907-06034}. The
DarkNetP authors divide models into only two partitions, one which runs in the Normal world and one which runs in
the Secure world. Our formulation expands this concept to treat each individual
layer as a partition with the ability to run arbitrary partitions in the Secure
world.  The use of branched execution partitioning is derived from the work of
Sutton et al.~\cite{5178994}. In this paper, the authors propose the use of
hidden layers which are not fully connected in order to increase the learning
speed. They found that models of this topology
outperform fully-partitioned and fully-unpartitioned networks even on
real-world vision tasks.
 \section{Conclusions and Future Work}
\label{sec:conclusion}

We characterized the confidential deep learning problem and identified the
challenges of using ARM TrustZone to execute deep learning models on untrusted
end-user devices. We sketched out three potential solutions for partitioning
models to address the inherent problems of limited secure memory in trusted
execution environments. While we have yet to fully explore model partitioning, there are a number of orthogonal research challenges that must
also be addressed before confidential deep learning can be deployed to a
production environment. We discuss just a few below. 

One challenge is establishing a root of trust. In our testing environment, no
protection is provided against malicious modifications to the OP-TEE Secure
world OS while it is in storage. To prevent such attacks, a hardware root of
trust, perhaps using a Trusted Platform Module, would need to be leveraged as
part of a trusted boot sequence. Trusted boot is achievable on modern smartphones, but it requires collaboration with the phone manufacturer.
Additionally, a method of key distribution and management would need to be
devised to protect the confidentiality of the deep learning model. Model
providers will need to distribute their models in an encrypted form, likely
using a secret key that is unique for every device-model pair to prevent one
key disclosure from comprising every model on every device. Further, to verify
the integrity of the trusted application a form of remote attestation would
need to be implemented, i.e.,  guaranteeing to the model
provider that the trusted application has not been modified and the model can
be decrypted safely. 

Another important challenge is preventing side-channel leakage. There are known side-channel
attacks for leaking private information from TEEs. Such attacks rely on
execution behavior that is both observable by the attacker and dependent on
private data, e.g., memory access patterns, execution time, storage access
patterns, and network accesses. The most promising defense against these attacks
is to design and implement algorithms whose executions are \emph{data
oblivious}, e.g., memory access patterns do not depend on private
data~\cite{ohrimenko2016oblivious}. The extent to which model partitioning
operations are vulnerable to these side-channels and how to best protect them
is an important question that we plan to investigate. Fortunately, there has
been some recent work in this area; Tople et al. recently detailed a
side-channel against DNNs on cloud servers using SGX that can infer encrypted
inputs to the model. However, their proposed defense is not a complete solution
for our problem as it doesn't protect other aspects of the model from leaking
(e.g., the weights)~\cite{tople2018privado}.

\balance
\bibliographystyle{ACM-Reference-Format}
\bibliography{bib/bib}

\end{document}